\documentclass[floatfix,aps,showpacs,nofootinbib,amssymb,amsmath]{revtex4}
\usepackage{graphicx}

\begin{document}

\title{Cosmological constraints on the Hu-Sawicki modified gravity scenario}

\author{Matteo Martinelli} \email{matteo.martinelli@roma1.infn.it}
\affiliation{Dipartimento di Fisica and Sezione INFN,
Universita' di Roma ``La Sapienza'', Ple Aldo Moro 2, 00185, Italy}
\author{Alessandro Melchiorri} \email{alessandro.melchiorri@roma1.infn.it}
\affiliation{Dipartimento di Fisica and Sezione INFN,
Universita' di Roma ``La Sapienza'', Ple Aldo Moro 2, 00185, Italy}
\author{Luca Amendola} \email{amendola@mporzio.it}
\affiliation{INAF/Osservatorio Astronomico di Roma, Viale Frascati 33, 0040,
Monteporzio Catone, (Roma), Italy}

\date{\today}

\begin{abstract}
In this  paper we place new constraints on a $f(R)$ modified gravity model
recently proposed by Hu and Sawicki. After checking that the Hu and Sawicki model
produces a viable cosmology, i.e. a matter dominated epoch followed
by a late-time acceleration, we constrain some of its parameters
by using recent observations from the UNION compilation of luminosity
distances of Supernovae type Ia, including complementary information
from Baryonic Acoustic Oscillations, Hubble
expansion, and age data. We found that the data considered is unable to place
significant constraints on the model parameters and we discuss the
impact of a different assumption of the background model in cosmic parameters
inference.
\end{abstract}


\maketitle

%%%%%%%%%%%%%%%%%%%%%%%%%%%%%%%%%%%%%%%%%%%%%%%%%%%%%%%%%%%%%%%%%%%%%%%%%
%%%%%%%%%%%%%%%%%%%%%%%%%%%%%%%%%%%%%%%%%%%%%%%%%%%%%%%%%%%%%%%%%%%%%%%%%
\section{Introduction}

The recent cosmological data from Cosmic Microwave Background Anisotropies,
galaxy surveys and luminosity distance of type Ia supernovae are all providing
supporting evidence for a dark energy component, responsable for more
than $70 \%$ of the total energy budget in our universe (see e.g. \cite{wmap5}).
Several candidates have been suggested for explaining this component,
as, for example, minimally coupled scalar fields (see e.q. \cite{copeland}
and references therein). However, it may also be
that the cosmological evidence for acceleration comes from the wrong assumption
of general relativity, i.e. that no dark component is present but actually
a modification to gravity is at work. In this respect, $f(R)$ theories seem to
provide a quite large number of viable models (for a recent review see
\cite{silvestri} and \cite{Caldwell:2009ix}).
A particular $f(R)$ model that evades solar system test has been proposed by Hu
and Sawicki (\cite{HS}, HS hereafter). The model has a modified Einstein-Hilbert
action:

\begin{equation}
S=\int{d^4x \sqrt{-g} \big[\frac{R+f(R)}{2k^2}+L_m\big]}
\label{action}
\end{equation}

\noindent  $L_m$ is the matter lagrangian, $k^2=8\pi G$ and

\begin{equation}
f(R)=-m^2\frac{c_1\big(\frac{R}{m^2}\big)^n}{1+c_2\big(\frac{R}{m^2}\big)^n}
\end{equation}

\noindent with $m^2=k^2\rho/3$ and $c_1$, $c_2$ and $n$ as free parameters.

As shown in \cite{HS} this model is able to reproduce the late time
accelerated universe but with distinctive deviations from a cosmological
constant. In this paper we investigate the cosmological viability of the HS
model in more detail. After a brief description of the model,
in the next section we will show that the HS model satisfies indeed the general conditions
presented by \cite{amendola} as a viable $f(R)$ model. In Sec. III we compare
the HS model with current data from SN-Ia luminosity distances from the UNION
catalog (\cite{union}), Baryonic Acoustic Oscillation data (\cite {bao})
and age constraints from the analysis of Simon, Verde and Jimenez
(\cite{hubble}). We show that the current data is
fully compatible with the HS model and that, unless a prior on the matter
density is used, the parameters of the model are unconstrained.
In particular, we analyze the impact of the HS model in the determination
of the current matter density. As we will show, assuming the HS model instead
of the standard cosmological model, could relax the constraints on the effective
matter density. A future incompatibility between the values of the matter density
$\Omega_M$ determined from different datasets and under the assumption of
the standard $\Lambda$CDM model could therefore provide an hint for
a modified gravity scenario.

\section{The Hu-Sawicky model}

Let us briefly review in this section the basic equations and results of the
HS model. Varying the action in Eq. 1 with respect to the metric $g^{\mu\nu}$ one obtains
the modified Einstein equations:

\begin{equation}
G_{\mu\nu}+f_RR_{\mu\nu}-\big(\frac{f}{2}-\Box f_R\big)g_{\mu\nu}-\nabla_\mu
\nabla_\nu f_R=k^2T_{\mu\nu}
\end{equation}

\noindent where $f_R=df/dR$ and $f_{RR}=d^2f/dR^2$ and assuming a flat FRW metric, the modified Friedmann equation:

\begin{equation}
H^2-f_R(HH'+H^2)+\frac{f}{6}+H^2f_{RR}R'=\frac{k^2\rho}{3}
\end{equation}

\noindent with $'\equiv d/dlna$ and $\rho$ the matter density at present time.\\
Defining the new variables $y_H=(H^2/m^2)-a^{-3}$ and $y_R=(R/m^2)-3a^{-3}$ the
Friedmann equations can be expanded in a system of two ordinary differential equations:

\begin{equation}
y'_H=\frac{y_R}{3}-4y_H
\end{equation}

\begin{equation}
y'_R=9a^{-3}-\frac{1}{y_H+a^{-3}}\frac{1}{m^2f_{RR}}\big[y_H-f_R\big(
\frac{y_R}{6}-y_H-\frac{a^{-3}}{2}\big)+\frac{f}{6m^2}\big]
\end{equation}

In order to compare the HS model with the cosmological constraints
usually derived under the assumption of dark energy, it is useful
to introduce an {\it effective} dark energy component with present energy density
$\tilde{\Omega}_x=1-\tilde{\Omega}_m$ and equation of state $w_x$, where
$\tilde{\Omega}_m$ is the effective matter energy density at present time.\\
Of course, in reality no dark energy component is present, the only
component present is matter and modified gravity gives the acceleration.
Considering the Friedmann equation:

\begin{equation}
\frac{H^2}{H_0^2}=\frac{\tilde{\Omega}_m}{a^3}+\tilde{\Omega}_x
e^{\int^1_a{da\frac{3[1+w_x(a)]}{a}}}
\end{equation}

\noindent the effective equation of state parameter $w_x$
for the dark energy component is given by

\begin{equation}
w_x=-1-\frac{1}{3}\frac{y'_H}{y_H}
\end{equation}

The free parameters $c_1$ and $c_2$  that appear in Eq. 2 can be expressed
in function of the effective density parameters by:

\begin{equation}
\frac{c_1}{c_2}\approx6\frac{\tilde{\Omega}_x}{\tilde{\Omega}_m}
\end{equation}

\begin{equation}
\frac{c_1}{c_2^2}=-\frac{f_{R_0}}{n}\big(\frac{12}{\tilde{\Omega}_m}-9\big)^{n+1}
\end{equation}

\begin{figure}[h!]
\centering
\includegraphics[width=11cm]{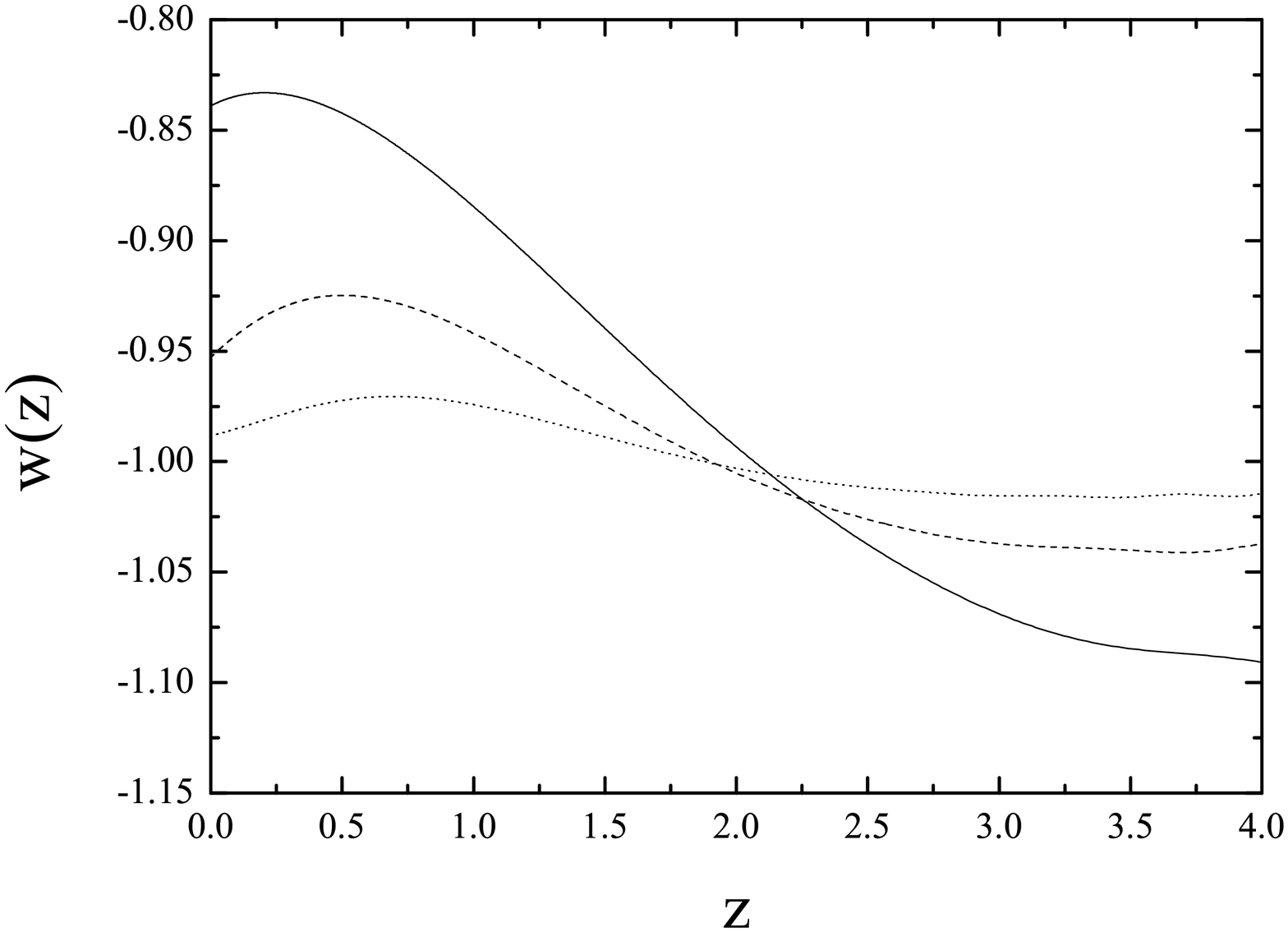}
\caption{The equation of state parameter $w_x$ for $n=1$ and $f_{R_0}$ equal
to $-0.1$ (solid line), $-0.03$ (dashed line) e $-0.01$ (dotted line). $\tilde{\Omega}_m$
is set to $0.24$}
\label{wn1}
\end{figure}

\begin{figure}[!h]
\centering
\includegraphics[width=11cm]{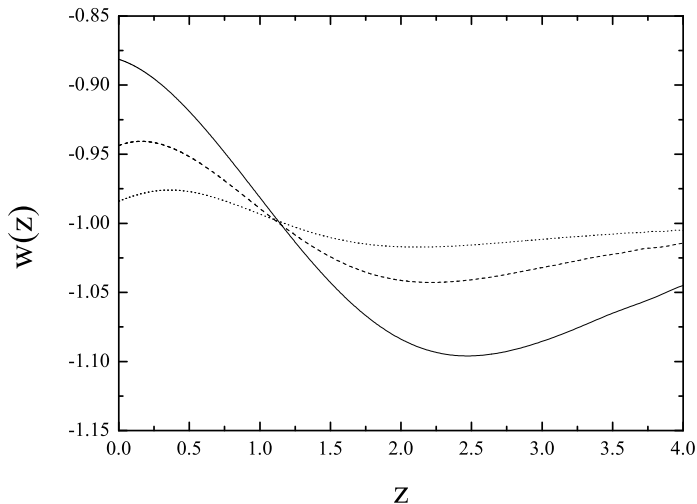}
\caption{The equation of state parameter $w_x$ for $n=2$ and $f_{R_0}$ equal
to $-0.1$ (solid line), $-0.03$ (dashed line) e $-0.01$ (dotted line). $\tilde{\Omega}_m$
is set to $0.24$}
\label{wn2}
\end{figure}

These relations show that the free parameters of the model are $\tilde{\Omega}_m$,
$n$, and $f_{R_0}$. The latter is constrained to $|f_{R_0}|\lesssim0.1$ by solar system tests \cite{HS}
and we will not investigate larger values in the next sections.\\
Solving the differential equations system for different values of $n$, $f_{R_0}$ and
$\tilde{\Omega}_m$, it is possible to obtain various evolution trends for the equation
of state parameter $w_x$. In Fig.\ref{wn1} and Fig.\ref{wn2} we plot the behavior of
$w_x$ in function of the redshift $z$ for different values of $f_{R_0}$ and $n$.\\
\\

As we can see in Fig.\ref{wn1} and Fig.\ref{wn2} the equation of state parameter $w_x$
follows a peculiar behavior in function of the redshift.
At the present time ($z=0$) $w_x$ has always a value higher than the
one predicted by the $\Lambda$CDM model
($w=-1$) and, moving towards higher redshifts, it decreases crossing
into the phantom region, i.e., assuming values lower than $-1$. For
even higher redshifts, $w_x$ moves asymptotically towards $-1$.\\
The same behavior is shown for any value of $n$ and $f_{R_0}$, moreover
decreasing the absolute value of $f_{R_0}$ brings $w_x$ closer to $-1$,
while increasing $n$ shifts the phantom crossing at lower redshift.

%\newpage

The modification of the Einstein-Hilbert lagrangian brings to a new equation
for the expansion of the Universe. The predicted expansion must be consistent
with standard cosmological results, i.e., should produce an accelerated era after
radiation and matter dominance.\\
Modified gravity models consistent with current observations, for example,
should not  change the scale factor evolution during
the matter era. Hence, it is possible to derive general conditions for
the cosmological viability of $f(R)$ theories.\\
Introducing the parameters

\begin{equation}
m(r)=\frac{Rf_{RR}}{1+f_R}\ \ \ r=-\frac{R(1+f_R)}{R+f}
\end{equation}

it is possible to show \cite{amendola} that for $f(R)$ theories the following
conditions apply:

\begin{itemize}
\item The model has a standard matter era with a following accelerated phase if

\begin{equation}
m(r)\approx0\ and\ m'(r)>-1\ \ \ with\ r\approx-1
\end{equation}

\item The accelerated phase goes asymptotically towards the one produced by a
dark energy with equation of state parameter $w=-1$, if

\begin{equation}
0\le m(r)\le1\ \ \ for\  r=-2
\end{equation}

\item The expansion is not of the phantom type ($w<-1$) if

\begin{equation}
m(r)=-1-r
\end{equation}

\end{itemize}

\begin{figure}[h!]
\centering
\includegraphics[width=11cm]{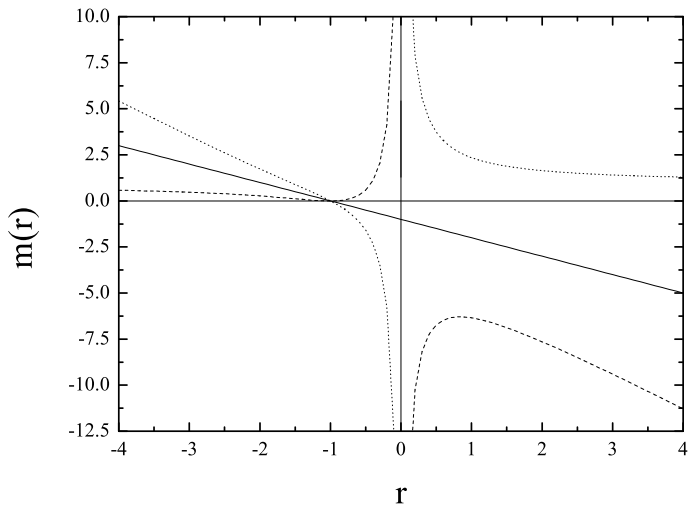}\caption{Plot of the two solution for
m(r) obtained setting $n=1$, $\tilde{\Omega}_m=0.3$ and $\tilde{\Omega}_\Lambda
=0.7$. The solid line corresponds to the viable solution while the dashed line
lies outside the viability region. The red line shows the solution $m(r)=
-r-1$.}
\label{m(r)}
\end{figure}

It is possible to calculate $m(r)$ for the Hu and Sawicki model and to show
the cosmological viability of this model.
In Fig.\ref{m(r)} we show that, for example, setting $n=1$,
$\tilde{\Omega}_m=0.3$ and $\tilde{\Omega}_\Lambda=0.7$, one obtains two
solutions for $m(r)$, one living outside the viability region and the other
corresponding to an acceptable expansion.

%%%%%%%%%%%%%%%%%%%%%%%%%%%%%%%%%%%%%%%%%%%%%%%%%%%%%%%%%%%%%%%%%%%%%%%%%
%%%%%%%%%%%%%%%%%%%%%%%%%%%%%%%%%%%%%%%%%%%%%%%%%%%%%%%%%%%%%%%%%%%%%%%%%

\section{Constraints on the HS model}
\label{secCMBanalysis}
%%%%%%%%%%%%%%%%%%%%%%%%%%%%%%%%%%%%%%%%%%%%%%%%%%%%%%%%%%%%%%%%%%%%%%%%%
%%%%%%%%%%%%%%%%%%%%%%%%%%%%%%%%%%%%%%%%
\subsection{Method}

In order to constrain the free parameters of the Hu and Sawicki model
($\tilde{\Omega}_m$, $n$ and $f_{R_0}$), we predicted the expected
theoretical values for a set of observables.\\
As now common in the literature, we considered the
 luminosity distance, defined by:

\begin{equation}
d_L(a)=\frac{1}{a}\int_a^1{\frac{da}{a^2H(a)}}=\frac{1}{aH_0}\int_a^1
{\frac{da}{a^2\sqrt{\tilde{\Omega}_m(y_H+a^{-3})}}}
\end{equation}

\noindent and the Hubble parameter:

\begin{equation}
H(a)=\sqrt{\tilde{\Omega}_mH_0^2(y_H+a^{-3})}
\end{equation}

Moreover, we also considered the quantity:

\begin{equation}
A\equiv\frac{\sqrt{\tilde{\Omega}_m}}{z_*}\big[z_*\frac{\Gamma^2(z_*)}
{\epsilon(z_*)}\big]^{\frac{1}{3}}
\end{equation}

\noindent where $z_*=0.35$, $\Gamma(z_*)=\int_0^{z_*}{dz/\epsilon(z)}$ and
$\epsilon(z)=H(z)/H_0$. The value of this parameter can be obtained from
observations of Baryon Acoustic Oscillations (BAO) \cite{constraints}.
Hence, we have another way to compare model prediction with data.\\

We used the superovae data from Kowalski et al. \cite{union} to obtain the
observational trend of $d_L(z)$ and we considered $H(z)$ values obtained by
Simon, Verde and Jimenez \cite{hubble} and a prior on the Hubble parameter
$H_0=0.72\pm0.08$ derived from measurements from the Hubble Space Telescope 
(HST, \cite{freedman}),
Finally, we used the value of $A$ from Eisenstein et al. \cite{bao}.\\

We compute a $\chi^2$ variable for each observational quantity and then
combine the results in a single variable 
$\chi^2=\chi^2_{SN}+\chi^2_{BAO}+\chi^2_{H}+\chi^2_{HST}$.
Once the theoretical evolution of the observational quantities is defined,
we can define a likelihood function as a function of $n$ and $f_{R_0}$ as

\begin{equation}
L=e^{-\frac{\chi^2-\chi^2_{min}}{2}}
\end{equation}

\noindent where $\chi^2_{min}$ is the minimum value in the considered range of $n$
and $f_{R_0}$.\\

\subsection{Results}

Combining the results obtained from the comparison between the experimental
data for $H(z)$, $A$ and $d_L(z)$ and their theoretical values, we can
constrain the free parameters $n$ and $f_{R_0}$ for different values of
$\tilde{\Omega}_m$ and $\tilde{\Omega}_x$.\\

\begin{figure}[h]
\centering
\includegraphics[width=12cm]{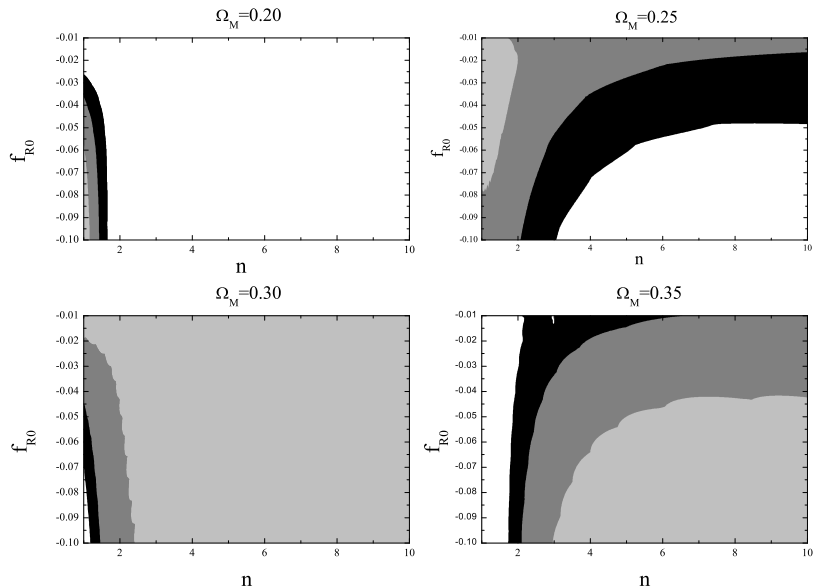}\caption{$68 \%$, $95 \%$ and $99 \%$ confidence
 levels in the $n$-$f_{R_0}$ plane in function of different values of $\tilde{\Omega}_m$.}
\label{chi37}
\end{figure}

Setting $\tilde{\Omega}_m=0.2$ and $\tilde{\Omega}_x=0.8$
it is possible to find an upper limit on $n$ and on $f_{R_0}$, $n<1.6$ and $f_{R_0}<-0.03$ at 2 $\sigma$,
while, performing the same analysis with different values of $\tilde
{\Omega}_m$ and $\tilde{\Omega}_\Lambda$, we can see that raising $\tilde
{\Omega}_m$ brings to more loosely constrained parameters.\\
We can note anyway that for higher values of $\tilde{\Omega}_m$, higher $n$
are preferred, while smaller values of $n$ are more in agreement with data
for smaller $\tilde{\Omega}_m$.\\
As we can see from Figure \ref{chi37}, for $\tilde{\Omega}_m=0.3$  both parameters are almost
totally unconstrained; this points out the need of an independent measurement of
the effective matter content ($\tilde{\Omega}_m$) in order to obtain some constraints
on $n$ and $f_{R_0}$.\\

It is interesting to compare the best fit values of $\chi^2$ obtained in the modified
gravity framework with the $\chi^2$ of a cosmological constant model. In order to quantify
the goodness-of-fit of the two models we use the Akaike information criterion (AIC) \cite{aic}
and the Bayesian information criterion (BIC) \cite{bic}, defined as
$$AIC=-2\ln L+2k$$
$$BIC=-2\ln L+k\ln N$$
where $L$ is the maximum likelihood, $k$ the number of parameters and $N$ the number of points.\\

\begin{figure}[h]
\centering
\includegraphics[width=9cm]{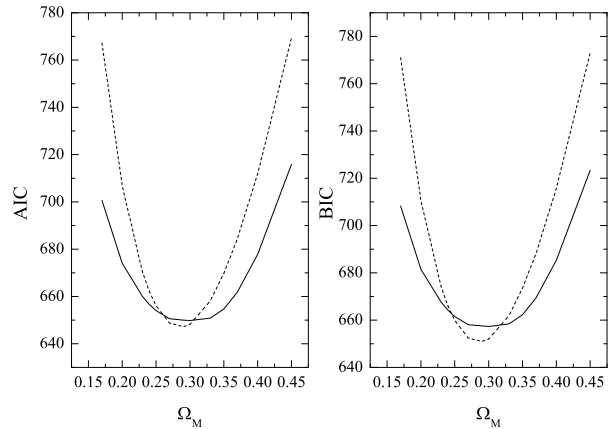}\caption{AIC (left panel) and BIC (right panel) tests
in function of $\Omega_m$ for the standard case of a cosmological constant (dashed line) and for
the HS model (solid line).}
\label{aic}
\end{figure}

In Fig. \ref{aic} we plot the best fit values of the $AIC$ and $BIC$ tests 
in function of  $\tilde{\Omega}_m$ for the standard model based on a cosmological constant and
the HS model respectively. As we can see, while the cosmological constant gives
slightly better values for the overall best fit, when larger or smaller values
of $\tilde{\Omega}_m$ are considered the $AIC$ and $BIC$ tests provide definitly better
values for the HS model. In few words, there is a weaker dependence of the 
 observables considered from $\tilde{\Omega}_m$ in the case of HS scenario.

\begin{figure}[h]
\centering
\includegraphics[width=13cm]{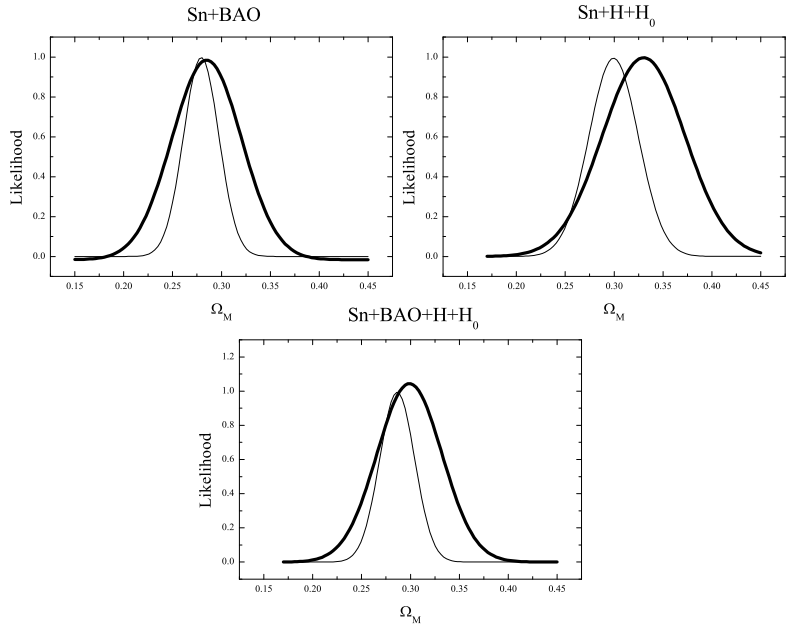}\caption{The likelihood function 
$L(\tilde{\Omega}_m)$ for the 
Hu and Sawicki model (thicker lines) and $\Lambda CDM$ model (fainter lines).}
\label{limitiom}
\end{figure}

It is therefore important to quantify the impact of a different choice of the theoretical
 background model on the derived constraints on $\tilde{\Omega}_m$. In Fig. \ref{limitiom} we compare the
constraints on the $\tilde{\Omega}_m$ parameter derived under the assumption of the HS
scenario with the similar constraints but assuming general relativity and dark energy.
As we can see the  $\tilde{\Omega}_m$ parameter is less constrained respect to
the $\Lambda$CDM scenario. This is certainly due to the larger amount of parameters
present in the HS model. In the future, with the increasing experimental accuracy,
if a discrepancy between independent constraints on the matter density will be found
 then a modified gravity scenario could be suggested as possible explanation.
This result anyway shows that one should also be extremely careful in considering the
current cosmological constraints because of their model dependence.

%%%%%%%%%%%%%%%%%%%%%%%%%%%%%%%%%%%%%%%%%%%%%%%%%%%%%%%%%%%%%%%%%%%%%%%%%
%%%%%%%%%%%%%%%%%%%%%%%%%%%%%%%%%%%%%%%%%%%%%%%%%%%%%%%%%%%%%%%%%%%%%%%%%

\section{Conclusions}
%%%%%%%%%%%%%%%%%%%%%%%%%%%%%%%%%%%%%%%%%%%%%%%%%%%%%%%%%%%%%%%%%%%%%%%%%
%%%%%%%%%%%%%%%%%%%%%%%%%%%%%%%%%%%%%%%%%%%%%%%%%%%%%%%%%%%%%%%%%%%%%%%%%

In this paper we have compared a modified gravity scenario, the HS model,
with several current cosmological datasets. We have found that the model
is in excellent agreement with recent SN-Ia, BAO and $H(z)$ data. Moreover,
the parameters of the model are substantially
unconstrained by the data considered. This has important effect on the
current constraints on some parameters as the matter density. We have shown
that the assumption of the HS model enlarges the current constraints on
this parameter by $\sim 30 \%$. If a discrepancy between two experimental
determinations of the matter density will be found in the framework of general
relativity, then a possible solution could be the introduction of a modified
gravity scenario. It will be duty of future experiments to scrutinize this
interesting possibility.

%%%%%%%%%%%%%%%%%%%%%%%%%%%%%%%%%
%%%%%%%%%%%%%%%%%%%%%%%%%%%%%%%%%%%%%%%%%%%%%%%%%%%%%%%%%%%%%%%%%%%%%%%%%
\acknowledgments
%%%%%%%%%%%%%%%%%%%%%%%%%%%%%%%%%%%%%%%%%%%%%%%%%%%%%%%%%%%%%%%%%%%%%%%%%
%%%%%%%%%%%%%%%%%%%%%%%%%%%%%%%%%%%%%%%%%%%%%%%%%%%%%%%%%%%%%%%%%%%%%%%%%

This research has been supported by ASI contract I/016/07/0 ``COFIS.''

%%%%%%%%%%%%%%%%%%%%%%%%%%%%%%%%%%%%%%%%%%%%%%%%%%%%%%%%%%%%%%%%%%%%%%%%%
%%%%%%%%%%%%%%%%%%%%%%%%%%%%%%%%%%%%%%%%%%%%%%%%%%%%%%%%%%%%%%%%%%%%%%%%%

%%%%%%%%%%%%%%%%%%%%%%%%%%%%%%%%%%%%%%%%%%%%%%%%%%%%%%%%%%%%%%%%%%%%%%%%%
%%%%%%%%%%%%%%%%%%%%%%%%%%%%%%%%%%%%%%%%%%%%%%%%%%%%%%%%%%%%%%%%%%%%%%%%%

\newpage

\end{document}